\documentclass[prb,twocolumn,showpacs,preprintnumbers,amsmath,amssymb]{revtex4}
\usepackage{graphicx}% Include figure files
\usepackage{dcolumn}% Align table columns on decimal point
\usepackage{caption2}
\usepackage{bm}% bold math

\begin{document}

\preprint{APS/123-QED}

\title{Von Neumann entropy and localization properties of two interacting
particles in one-dimensional nonuniform systems}
\author{Longyan Gong  $^{1,2,3}$}
\thanks{Email address:lygong@njupt.edu.cn.}
\author{Peiqing Tong  $^{2,}$ }
\thanks{Corresponding author. Email address:pqtong@njnu.edu.cn.}
\affiliation{
 $^{1}$Department of Mathematics, Nanjing Normal University,
Nanjing, Jiangsu, 210097,P.R.China\\
 $^{2}$Department of Physics, Nanjing Normal University,
Nanjing, Jiangsu, 210097,P.R.China\\
 $^{3}$Department of Mathematics and
Physics, Nanjing University of Posts and Telecommunications,
Nanjing, Jiangsu 210003, P.R.China
}%
\date{today}
\begin{abstract}
With the help of von Neumann entropy, we study numerically the
localization properties for two interacting particles (TIP) with
the on-site interaction in one-dimensional disordered,
quasiperiodic and slowly varying potential systems, respectively.
We find that for TIP in disordered and slowly varying potential
systems, the spectrum-averaged von Neunmann entropy $\langle
E_v\rangle$ first increases with interaction $U$ until its peaks,
then decreases as $U$ gets larger.  For TIP in the Harper model,
the functions of $\langle E_v\rangle$ versus $U$ are different for
particles in extended and localized regimes. Our numerical results
indicate that for these two-particle systems, the von Neumann
entropy is a suitable quantity to characterize the localization
properties of particle states. Moreover, our studies propose a
consistent interpretation of the discrepancies between previous
numerical results.

\end{abstract}
\maketitle

\section{Introduction}
Interacting uniform electronic systems\cite{li68} and
non-interacting  disordered electronic systems\cite{an58,kr93} are
two of the most intriguing, albeit difficult, subjects in
condensed matter physics.  As a result of the complexity of the
simultaneous presence of randomness and interactions,  few
definitive results are known.\cite{le85,be94} To understand the
effects of electrons interaction on the localization properties in
a random potential, Shepelyansky proposed some years earlier that
it would be worthwhile to consider the simple case of two
interacting particles(TIP) in a one-dimensional (1D) random
potential.\cite{sh94} Subsequently, extensive efforts have been
devoted to TIP in various systems.\cite{sh96,ev97,br06,ev96,op96,
so97,ro97} In particular, two interacting electrons with total
spin zero in a 1D Harper model,\cite{ev97} in Fibonacci and
Thue-Morse lattices \cite{br06} are studied. The behavior of TIP
has been studied using the time evolution of wave packets
,\cite{sh94,sh96,ev97,br06} exact diagonalization
,\cite{ev97,ev96} Green function \cite{op96, so97,fr97} and
transfer-matrix,\cite{ro97} $etc$. Due to different definitions of
the localization length for TIP and different methods applied,
there are discrepancies about the effect of the combination of
disorder and interaction. For example, for TIP in 1D disordered
potentials, Shepelyansky \cite{sh94} and others
 \cite{op96,so97} have found that for small interaction strengths,
the two-particle interaction can enhance the Anderson localization
length, while R\"{o}mer and Schreiber \cite{ro97} found  no
enhancement of the localization length when the system size grows
to infinity. Evangelou $et$ $al.$ have pointed out that stronger
localization occurs at large interaction when compared to the
noninteracting case.\cite{ev96} At the same time, for TIP in a 1D
Harper model, it was found that the interaction would induce
localization effect,\cite{sh96,ba96,ba97} while Evangelou and
Katsanos \cite{ev97} found that the effects of particle
interactions are different for electrons in extended and localized
regimes.

On the other hand,  quantum entanglement, which attracting much
attention in quantum information,\cite{bo00} has been extensively
applied in condensed matter
physics.\cite{za02,yu04,gu04,la06,go06,bu06} For examples, quantum
entanglement measured by the von Neumann entropy was studied in
the Hubbard model for the dimer case \cite{za02}, in the extended
Hubbard model for different band filling\cite{gu04}, in quantum
small-world networks\cite{go06}, and in low-dimensional
semiconductor systems \cite{bu06}.  It was found that the von
Neumann entropy is suitable for analyzing the interplay between
itinerant and localized features \cite{za02}, as well as
characterizing quantum phase transition\cite{gu04,la06} and the
localization-delocalization transition of electron
states\cite{go06}.

In this paper, we perform a detailed study of the von Neumann
entropy for TIP in 1D disordered and quasiperiodic systems
respectively, taking into account different on-site interactions
$U$ at various on-site potential strengths. We find that for TIP
in disordered and slowly varying potential systems, the
spectrum-averaged von Neunmann entropy $\langle E_v\rangle$ first
increases with interaction $U$ until its peaks, then decreases as
$U$ gets larger. For TIP in the 1D Harper model, the functions of
$\langle E_v\rangle$ versus $U$ are different for particles in
extended and localized regimes. Finally we study a two-particle
system based on the slowly varying potential model
.\cite{gr88,th88,sa88} From these studies, we can conclude that
for TIP systems, the von Neumann entropy is a suitable quantity to
characterize the localization properties of particle states.

The paper is organized as follows. In the next section the von
Neumann entropy is described. In Sec.~\ref{sec3} the numerical
results for TIP in 1D disordered, quasiperiodic, and slowly
varying potential systems are presented, respectively. And we
present our conclusions and discussions in Section ~\ref{sec4}.

\section{\label{sec2}TIP models and von Neumann entropy}
\subsection{\label{sec21}TIP models}
Following recent literature\cite{sh94,sh96,so97,ro97,ba96,ba97}
the eigenvalue equation for the TIP in 1D system can be written
as
\begin{eqnarray}
(\varepsilon_{n_1}+\varepsilon_{n_2}+U\delta_{n_1,n_2})\psi_{n_1,n_2}
 +t(\psi _{{n_1+1},n_2}\nonumber\\
+\psi_{n_1-1,n_2} +\psi_{n_1,n_2+1}+ \psi_{n_1,
n_2-1})=E\psi_{n_1,n_2}, \label{form1}
\end{eqnarray} %formu(1)
where $\varepsilon_n$ is the on-site potential, $t$ is a
nearest-neighbor hopping integral and $U$ characterizes the
on-site interaction between particles.

Eq.(\ref{form1}) actually can  describe the behaviors both bosons
and fermions, i.e., two spinless bosons or two electrons with
opposite spins.

For spinless bosons with the on-site interaction, the
tight-binding Hamiltonian can generally be described by
\begin{eqnarray}
&&H =t\sum\limits_{n = 1}^{N-1} {(c_n^ \dag  c_{n + 1}  + c_{n +
1}^ \dag c_n)}
+\sum\limits_{n = 1}^N {\varepsilon_n c_n^ \dag c_n}\nonumber\\
&&+U\sum\limits_{n = 1}^N {(c_n^ \dag  c_n)(c_n^ \dag c_n)},
\label{form2}
\end{eqnarray} %formu(2)
where $c_n^\dag$ ($c_n$) is the boson creation(annihilation)
operator of the \emph{n}th site. The generic eigenstate for two
spinless bosons is the superposition
\begin{equation}
\left| \alpha  \right\rangle  = \sum\limits_{n_1\leq n_2}^N {\psi
^\alpha_{n_1, n_2} } \left| n_1, n_2 \right\rangle =
\sum\limits_{n_1\leq n_2}^N {\psi^\alpha _{n_1, n_2} c_{n_1}^ \dag
c_{n_2}^ \dag } \left| 0 \right\rangle, \label{form3}
\end{equation} %formu(3)
where $\left| 0 \right\rangle$ is the vacuum and
${\psi^\alpha_{n_1, n_2} }$ is the amplitude of wave
function. %Here ${\psi^\alpha_{n_1, n_2} }={\psi^\alpha_{n_2, n_1} }$. %For
%simplicity we restrict that $n_1\leq n_2$ and the normalization
%condition is $ \sum\limits_{n_1,n_2(n_1\leq n_2)}^N \left
%|\psi^{B\alpha}_{n_1, n_2} \right | ^2=1$.
From Eqs.(\ref{form2}) and (\ref{form3}) we can obtain the
eigenvalue equation (\ref{form1}).

For electrons with the on-site interaction, the tight-binding
Hamiltonian can be described by \cite{ev97,ev96,br06}
\begin{eqnarray}
&&H =t\sum\limits_{n = 1}^N \sum\limits_{\sigma}{(c_{n,\sigma}^
\dag c_{n + 1,\sigma}  + c_{n + 1,\sigma}^ \dag
c_{n,\sigma})}\nonumber\\
&&+\sum\limits_{n = 1}^N\sum\limits_{\sigma} {\varepsilon_n
c_{n,\sigma}^ \dag c_{n,\sigma}} +U\sum\limits_{n = 1}^N
{c_{n,\uparrow}^ \dag c_{n,\uparrow}c_{n,\downarrow}^ \dag
c_{n,\downarrow}}, \label{form4}
\end{eqnarray} %formu(4)
where $c_{n,\sigma}^\dag$ ($c_{n,\sigma}$) is the electron
creation(annihilation) operator for the electron at the \emph{n}th
site with spin $\sigma=\pm\frac{1}{2}$.  For two electrons the
Hilbert space can be conveniently divided into a singlet subspace
with total spin $S=0$ and a triplet subspace with total spin
$S=1$, respectively. Since the triplet subspace permits no double
occupation, it is not affected by the on-site interaction. In
order to analyze the effect of the on-site interaction $U$, we
will consider the case of the two electrons with opposite spins,
i.e., the singlet subspace. In a chain with $N$ sites the singlet
subspace is spanned by $N(N + 1)/2$ spatially symmetric basis
functions
\begin{eqnarray}
 \lefteqn{\left| n_1, n_2 \right\rangle} \hspace{8cm}
 \nonumber\\
=\left\{
\begin{array}
{r@{\quad\quad}l} \frac{1}{\sqrt{2}}(c_{n_1,\uparrow}^\dag
c_{n_2,\downarrow}^\dag+c_{n_2,\uparrow}^\dag
c_{n_1,\downarrow}^\dag)\left| 0 \right\rangle $ $ \textrm{for} $
$ n_1<n_2,
\\(c_{n_1,\uparrow}^\dag
c_{n_2,\downarrow}^\dag)\left| 0 \right\rangle $ $ \textrm{for} $
$ n_1=n_2,
\end{array}\right.\label{form5}
\end{eqnarray} %form5
where $\left| 0 \right\rangle$ is the vacuum. An eigenstate for
two electrons with the spatially symmetric wave functions is in
general the superposition
\begin{eqnarray}
&&\left| \alpha  \right\rangle =\sum\limits_{n_1\leq n_2}^N {\phi
^\alpha_{n_1, n_2} } \left| n_1, n_2 \right\rangle, \label{form6}
\end{eqnarray} %formu(6)
where  $\phi ^\alpha_{n_1, n_2}$ is the amplitude of wave
function. By making the transformation
\begin{eqnarray}
 \psi ^{\alpha}_{n_1, n_2}
=\left\{
\begin{array}
{r@{\quad\quad}l} \phi ^{\alpha}_{n_1, n_2} $ $ \textrm{for} $ $
n_1<n_2,\nonumber\\
\\\sqrt {2}\phi
^{\alpha}_{n_1, n_2} $ $ \textrm{for} $ $ n_1=n_2,
\end{array}\right.
\end{eqnarray}
the eigenvalue equation obtained from Eqs.
(\ref{form4}---\ref{form6}) can be written as equation
(\ref{form1}).

In the following our numerical method is described for bosons. The
extension for fermions is straightforward.

\subsection{\label{sec22}von Neumann entropy}
For the two particles in the system we are studying, there are
three local states at each site, $\left| 2 \right\rangle_n, \left|
1 \right\rangle_n, \left| 0 \right\rangle_n$, corresponding to the
state with two, one or zero particles at the \emph{n}th site,
respectively. The local density matrix $\rho_n$ is defined
\cite{za02,gu04} by
\begin{eqnarray}
\rho_n&=& z_{2n}\left| {2}\right\rangle{_n}{_n}\left\langle{2}
\right|+z_{1n}\left| {1}\right\rangle{_n}{_n}\left\langle{1}
\right| \nonumber\\
&+&(1-z_{1n}-z_{2n})\left|{0}\right\rangle{_n}{_n}\left\langle{0}
\right|.\label{form7}
\end{eqnarray} %formu(7)
For two spinless bosons,
\begin{eqnarray}
z_{2n}=\left\langle \alpha \right|c_n^ \dag  c_n c_n^ \dag
c_n\left| \alpha \right\rangle=\psi^{\alpha} _{n,n}\psi^{*\alpha}
_{n,n}\label{form8}
\end{eqnarray}%formu(8)
and
\begin{eqnarray}
 &z_{1n}&=\left\langle \alpha \right|c_n^ \dag c_n \left| \alpha
\right\rangle-2z_{2n} \nonumber\\
&=&\sum\limits_{m(m>n)}^N \psi^{\alpha} _{n,m}\psi^{*\alpha}
_{n,m}+\sum\limits_{m'(m'<n)}^N \psi^{\alpha}
_{m',n}\psi^{*\alpha} _{m',n}.\label{form9}
\end{eqnarray}%formu(9)
Consequently, the corresponding von Neumann entropy related to
$n$th site is

\begin{eqnarray}
E^\alpha_{vn}=-(1-z_{1n}-z_{2n})\log_2(1-z_{1n}-z_{2n})
\nonumber\\-z_{1n}\log_2z_{1n}-z_{2n}\log_2z_{2n}.\label{form10}
\end{eqnarray}%formu(10)
For nonuniform systems, the value of $E^\alpha_{vn}$ depends on
the site position $n$. At an eigenstate $\alpha$, we define a
site-averaged von Neumann entropy
\begin{equation}
E^\alpha_v= \frac{1}{N} \sum\limits_{n=1}^N
{E^\alpha_{vn}}.\label{form11}
\end{equation} %formu11
For a delocalized state that all
$\psi^{\alpha}_{n_1,n_2}=\frac{1}{\sqrt{N(N+1)/2}}$ for all
$n_1\leq n_2$, this definition gives $E^\alpha_v \approx
\frac{2}{N}\log_2{\frac{N}{2}}$ at large $N$, while for a
localized state that $\psi^{\alpha}_{n_1^\circ,n_2^\circ}=1$ at
given $n_1^ \circ$ and $n_2^ \circ$, $E^\alpha_v=0$. In this
paper, all the values of $E^\alpha_v$ are scaled by
$\frac{2}{N}\log_2{\frac{N}{2}}$. From the two examples, we know
that the scaled $E^\alpha_v$ is close to $1$ for eigenstates are
extended and almost vanishes for eigenstates are localized.
Henceforth, we omit ``scaled" for simplicity.

In order to analyze the influence of system parameters like the
on-site interaction $U$,  on the von Neumann entropy for all the
eigenstates, we define a spectrum-averaged von Neumann entropy as
a further gross measure
\begin{equation}
\langle E_v \rangle  = \frac{1}{M}\sum\limits_{\alpha} {E^\alpha_v
},\label{form12}
\end{equation} %formu(12)
where $M$ is the number of all the eigenstates.

\section{\label{sec3} numerical results}
From now we consider only the repulsive interaction ($U>0$). We
directly diagonalize the eigenvalue Eq.(\ref{form1}) with the
periodic boundary condition at finite system sizes and obtain all
eigenvalues $E_\alpha$ and the corresponding eigenvectors $\left|
\alpha \right\rangle$. Without loss of generality,  the hopping
integral $t$ is taken as units of energy. From the formulas
(\ref{form7}---\ref{form12}), we then can obtain the site-averaged
von Neumann entropy $E^\alpha_v$ and the spectrum-averaged von
Neumann entropy $\langle E_v \rangle$, respectively.

\subsection{TIP in a disordered potential chain}
For TIP in a disordered potential chain, the on-site potential
$\varepsilon_n$ in Eq.(\ref{form3}) are random variables uniformly
distributed with $[-W,W]$. Here $W$ characterizes the degree of
on-site disorder as in the Anderson model.\cite{an58} For this
model, we calculate the spectrum-averaged von Neumann entropy
$\langle E_v \rangle$ with  a given set of  parameters $W$ and
$U$. For every set of parameters $W$ and $U$, the disorder average
is taken over $100$ samples.\cite{go07} More samples simply give
similar results.

%----------------------------------------
\begin{figure}%[!ht]%fig1
\includegraphics[width=2.5in]{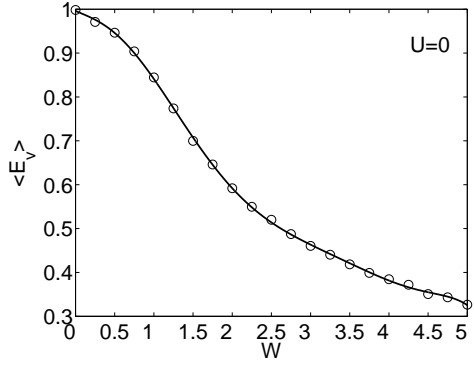}
\caption{ The spectrum-averaged von Neumann entropy $\langle
E_v\rangle$ versus the disorder $W$ for $U=0$ at
$N=90$.}\label{Fig1}
\end{figure} %fig1

\begin{figure}%[!ht]%Fig2
\includegraphics[width=2.5in]{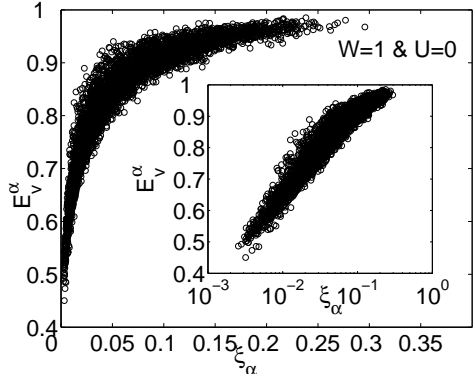}
\caption{The relation between the site-averaged von Neumann
entropy $E^\alpha_v$ and the corresponding $\xi_\alpha$ at $W=1$
and $U=0$ for a typical sample. The $E^\alpha_v$ versus
$\log_{10}\xi_\alpha$ is plotted in the inset.}\label{Fig2}
\end{figure} %fig2

For 1D Anderson model in the absence of the interaction($U=0$), it
is well known that all the eigenstates are localized and the
one-particle localization length is $\xi\approx25 t^2/W^2$ at the
energy band center.\cite{kr93} Fig.\ref{Fig1} gives the
spectrum-averaged von Neumann entropy $\langle E_v\rangle$ versus
the disorder parameter $W$. It shows that  $\langle E_v\rangle$
monotonically decreases as $W$ increases, reflecting the trivial
localization effect of the on-site disorder in the model. To find
the correlation between $\langle E_v\rangle$ and the localization
properties for TIP systems, we also study the inverse
participation ratio (IPR)\cite{ba96}, defined by $\xi_\alpha  =
(\frac{N(N+1)}{2}\sum\limits_{n_1\leq n_2}^{N} {|\psi _{n_1 ,n_2
}^ \alpha|^4 )^{- 1}}$, which gives the ratio of lattice sites
occupied by particles to all lattice sites at an eigenstate
$\alpha$. The larger $\xi_\alpha$ is, the more delocalized the
 eigenstate is. %Though there
In Fig.\ref{Fig2} we plot the site-averaged von Neumann entropy
$E^\alpha_v$ versus $\xi_\alpha$ at $U=0$ for a typical sample
with $W=1$. On the whole, $E^\alpha_v$ increases logarithmically
with $\xi_\alpha$ as can been seen in the inset of Fig.\ref{Fig2},
so von Neumann entropy can well reflect the localization
properties of two-particle eigenstates.

\begin{figure}%[!ht]%Fig3
(a)\includegraphics[width=2.5in]{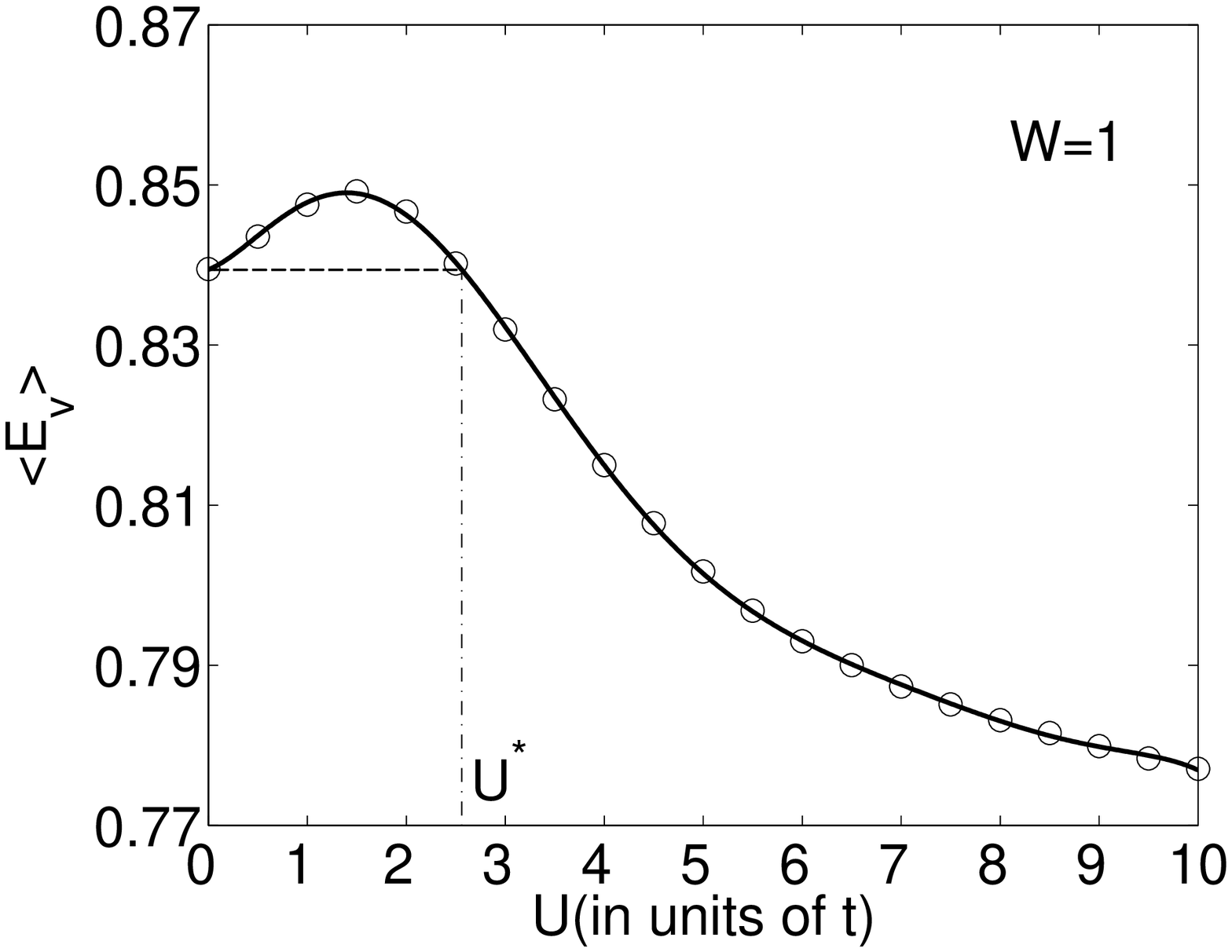}
(b)\includegraphics[width=2.5in]{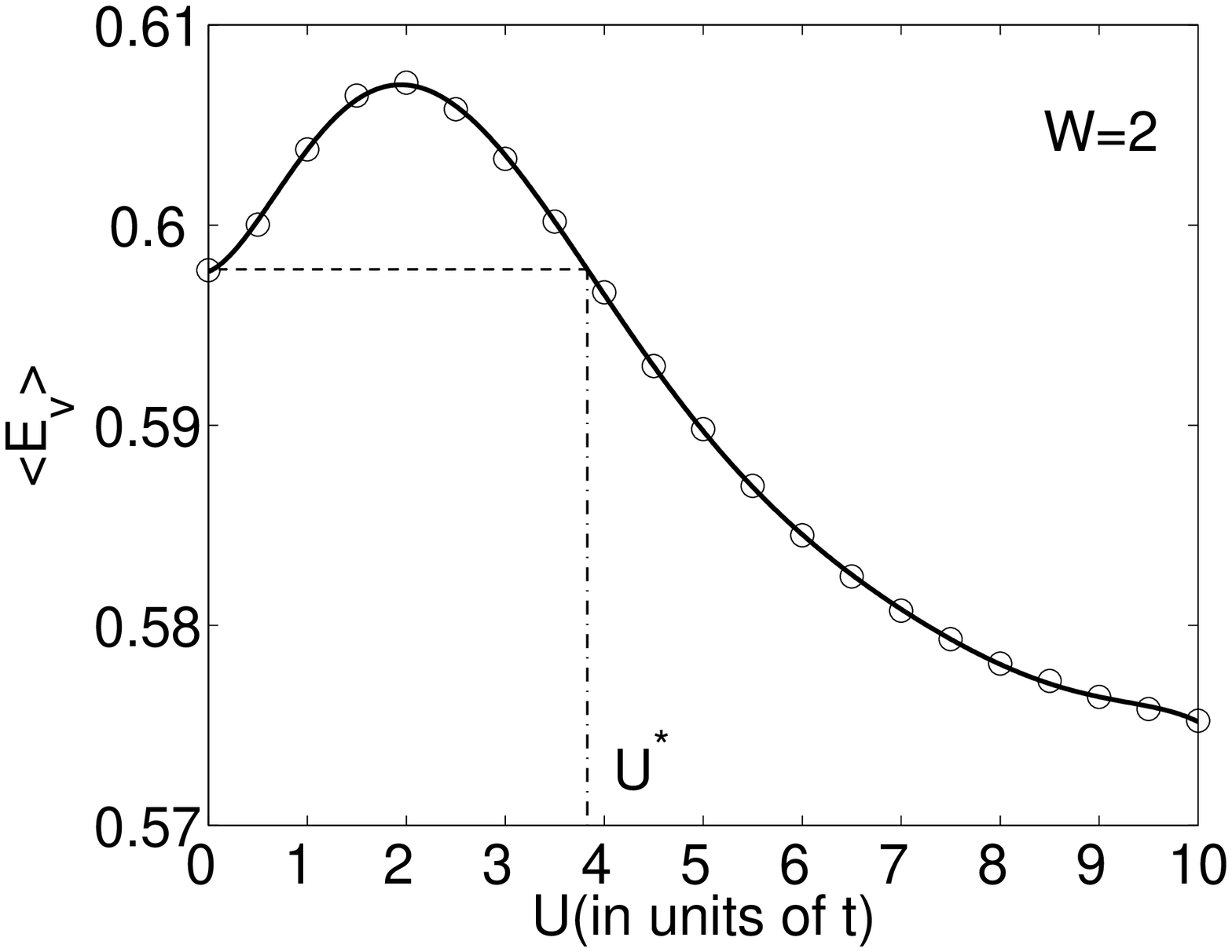}
(c)\includegraphics[width=2.5in]{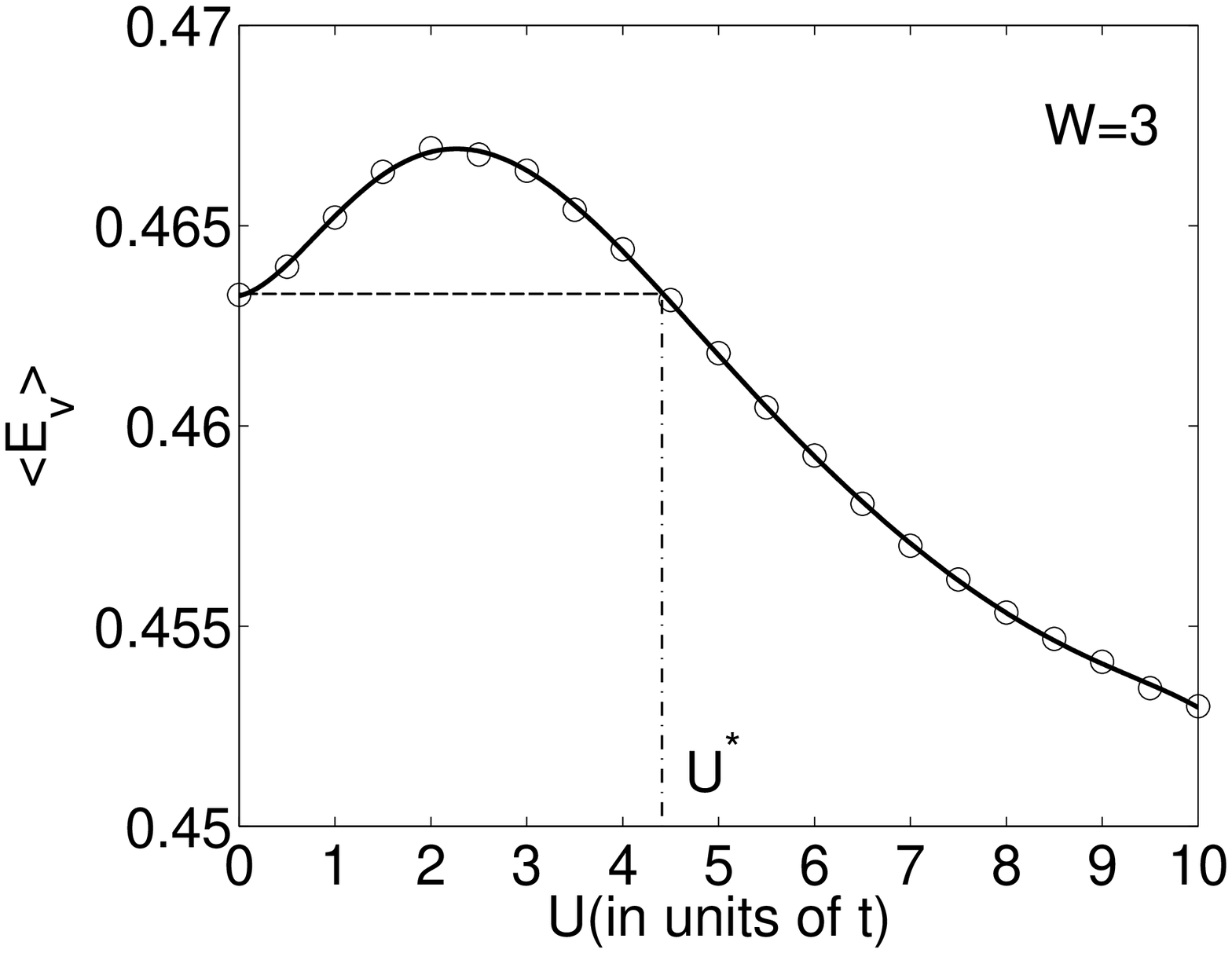}

\caption{ The spectrum-averaged von Neumann entropy $\langle
E_v\rangle$ as a function of interaction strengths $U$ for
(a)$W=1$,(b)$W=2$ and (c)$W=3$, respectively. Here
$N=90$.}\label{Fig3}
\end{figure} %Fig3

\begin{figure}%[!ht]%Fig4
\includegraphics[width=2.5in]{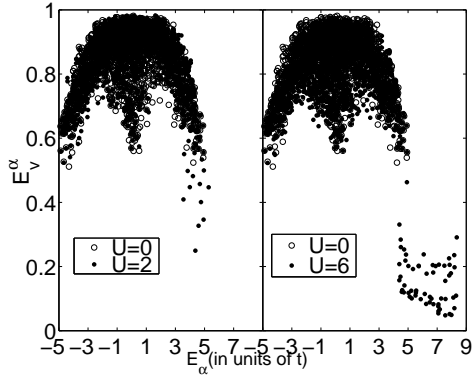}
\caption{The site-averaged von Neumann entropy $E^\alpha_v$ and
the corresponding eigenenergy $E_\alpha$ at different $U$ for a
typical sample at $W=1$. Here $N=90$ and the spectrum-averaged von
Neumann entropy $\langle E_v\rangle=0.8431,0.8556$ and $0.7916$
for $U=0,2$ and $6$, respectively.}\label{Fig4}
\end{figure}

For $U>0$, the spectrum-averaged von Neunmann entropy $\langle
E_v\rangle$ as a function  of $U$ at  $W=1,2,3$ are plotted in
Fig.\ref{Fig3}.  The results are similar for other $W$. For all
$W$, we find that $\langle E_v\rangle$ first increases until its
peak as $U$ increases from zero, then decreases as $U$ gets
larger. There is a $U^*$, at which $\langle E_v\rangle$ is equal
to that at $U=0$. When $U$ is smaller(greater) than $U^*$,
$\langle E_v\rangle$ is larger(smaller) than the value of $\langle
E_v\rangle$ at $U=0$. To understand the effect of the interaction
$U$ on the von Neumann entropy, we calculate the site-averaged von
Neumann entropy $E^\alpha_v$ and illustrate the results in
Fig.\ref{Fig4} at different $U$ for $W=1$. It shows that only a
small portion of the two-particle eigenstates have its value of
$E_v^\alpha$ changed by the interaction $U$. For small $U$, most
of the newly created eigenstates are in the main band and the
corresponding $E_v^\alpha$ are larger than that in the
noninteracting case, while for large $U$, most of them are above
the top of the main band and the corresponding $E_v^\alpha$ are
smaller than that for particles without interaction. Therefore,
the varying $\langle E_v\rangle$ with $U$ is not monotonic as
shown in Fig.\ref{Fig3}. These two competing effects reach an
equilibrium at $U=U^*$.

 %fig4

For TIP in 1D disordered potentials with $U\leq t$ Shepelyansky
\cite{sh94} and others \cite{op96,so97} have found that the
Anderson localization length increases with $U$ . It agrees with
our conclusions that for small $U$, $\langle E_v\rangle$ increases
with $U$. For large interaction $U$, Evangelou $et$ $al.$
\cite{ev96} have pointed out that stronger localization occurs
when compared to the noninteracting case, because at $U>>t$
interaction can significantly modify the energy spectrum
\cite{sh94} and the created eigenstates have the comparatively
small localization lengths.\cite{ev96} The result is consistent
with ours that for $U>U^*$, the values of $\langle E_v\rangle$ are
smaller than that at $U=0$.
%----------------------------------------

\subsection{TIP in a quasiperiodic potential chain}
From Eq.(\ref{form1}), the eigenvalue equation for TIP in a
quasiperiodic potential chain based on the Harper model can be
described by \cite{sh96,ev97}
\begin{eqnarray}
\lefteqn{[\lambda\cos(2\pi\sigma
n_1+{\beta}_1)+\lambda\cos(2\pi\sigma
n_2+{\beta}_2)+U\delta_{n_1,n_2}]\psi_{n_1,n_2} }\hspace{1cm}\nonumber\\
& &+t(\psi_{n_1+1, n_2} +\psi_{n_1-1, n_2} +\psi_{n_1, n_2+1}+ \psi_{n_1, n_2-1}) \nonumber\\
& &=E\psi_{n_1, n_2}, \label{form13}
\end{eqnarray} %formu(13)
here the parameter $\lambda$ characterizes the strength of the
quasiperiodic potential, $\sigma$ and $\beta_{1,2}$ are constants.
As a typical case, we set $\sigma=(\sqrt{5}-1)/2$,
$\beta_1=\beta_2=0$. As is customary in the context of
quasiperiodic system, the value of $\sigma$ may in fact be
approximated by the ratio of successive Fibonacci
numbers---$F_n=F_{n-2}+F_{n-1}$ . In this way, choosing
$\sigma=F_{n-1}/F_n\approx(\sqrt{5}-1)/2$ and the system size
$N=F_n$, we can obtain the periodic approximant for the
quasiperiodic potential. In our calculation, $N$ is chosen as
Fibonacci numbers $34, 55$ and $89$, respectively.

\begin{figure}%[!ht]%Fig5
(a)\includegraphics[width=2.5in]{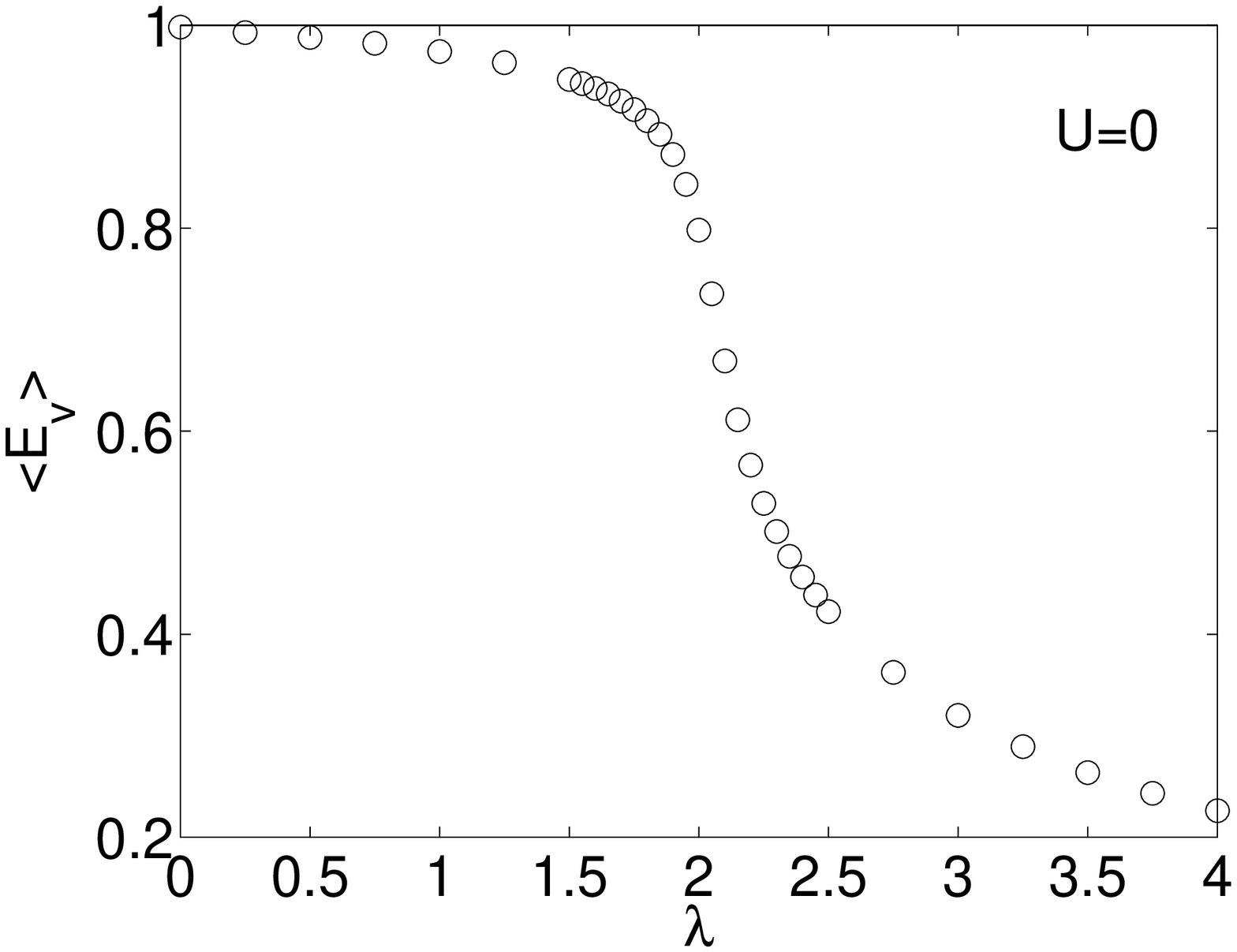}
(b)\includegraphics[width=2.5in]{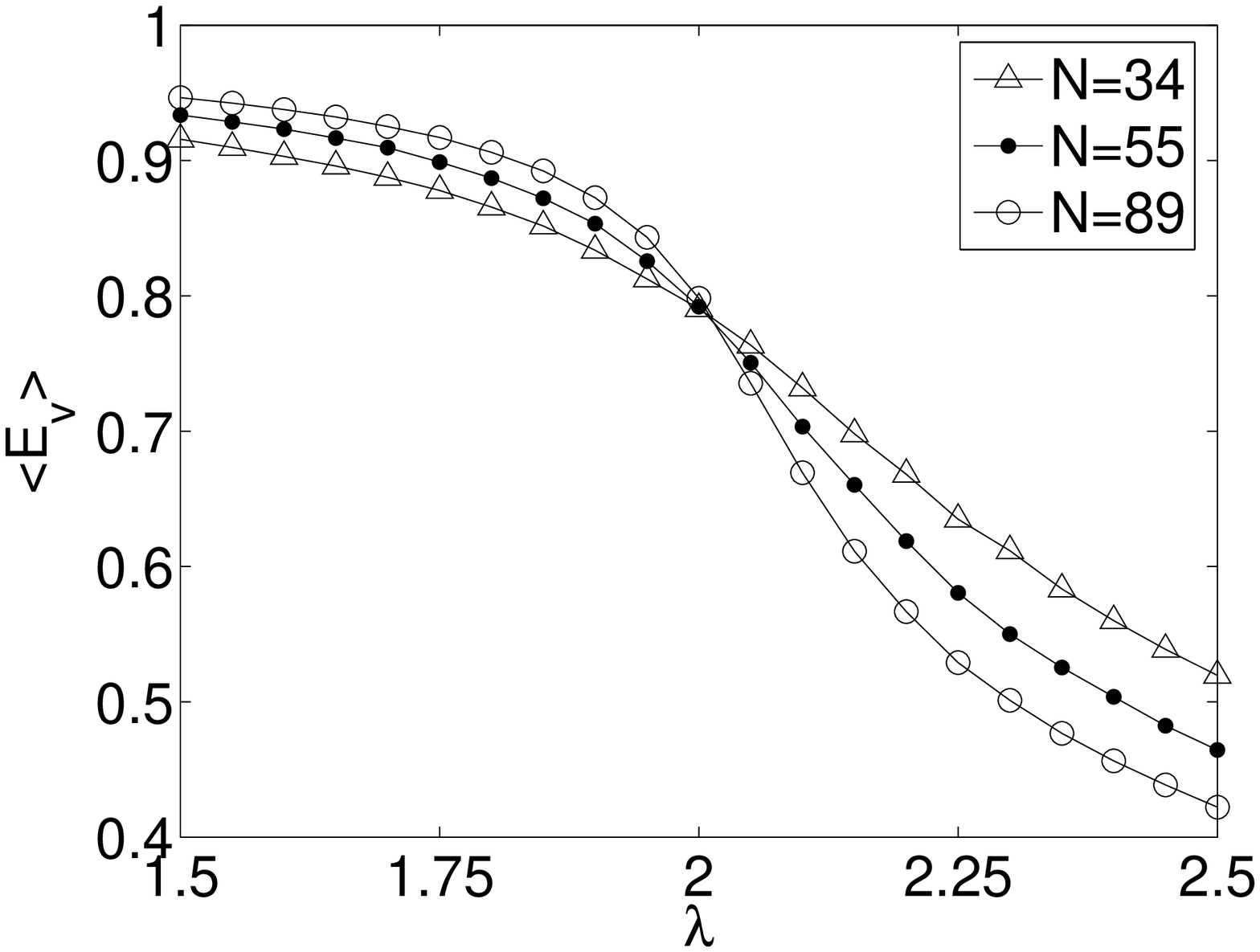}
 \caption{ (a)The
spectrum-averaged von Neumann entropy $\langle E_v\rangle$ as a
function of quasiperiodic potential strength $\lambda$ at $U=0$.
Here $N=89$. (b)The $\langle E_v\rangle$ versus $\lambda$ at $U=0$
for different $N$. }\label{Fig5}
\end{figure} %Fig5

In the absence of the interaction $U$ it is found \cite{au80,so85}
that that for $\lambda <2$ the spectrum becomes continues and all
eigenstates are extended. For $\lambda
>2$ the spectrum is pure point and all eigenstates are
exponentially localized.  For $\lambda =2$ the situation is
critical with a singular-continuous multifractal spectrum and
power law localized eigenstates. Metal-insulator transition can
occur at $\lambda=2$. Obviously they are different from these for
the 1D Anderson model where all eigenstates are localized for
$W\neq0$.

Fig.\ref{Fig5}(a) shows the spectrum-averaged von Neumann entropy
$\langle E_v\rangle$ with respect to the quasiperiodic potential
strength $\lambda$ at $U=0$. We observe that there is a sharp
decrease of $\left\langle {E_v} \right\rangle$ near the critical
value $\lambda_c=2$.  For $\lambda<\lambda_c$ which corresponds to
extended states, all $\left\langle {E_v} \right\rangle$ are around
$1$, while for $\lambda>\lambda_c$ corresponding to localized
states, all $\left\langle {E_v} \right\rangle$ are far less than
$1$. In Fig.\ref{Fig5}(b) we plot $\left\langle {E_v}
\right\rangle$ as a function of $\lambda$ at $U=0$ for $N=34,55$
and $89$. The curves cross at $\lambda\approx\lambda_c$. The
crossing point separates the extended($\lambda<\lambda_c$) and the
localized($\lambda>\lambda_c$)regimes.
\begin{figure}%[!ht]%Fig6
(a)\includegraphics[width=2.5in]{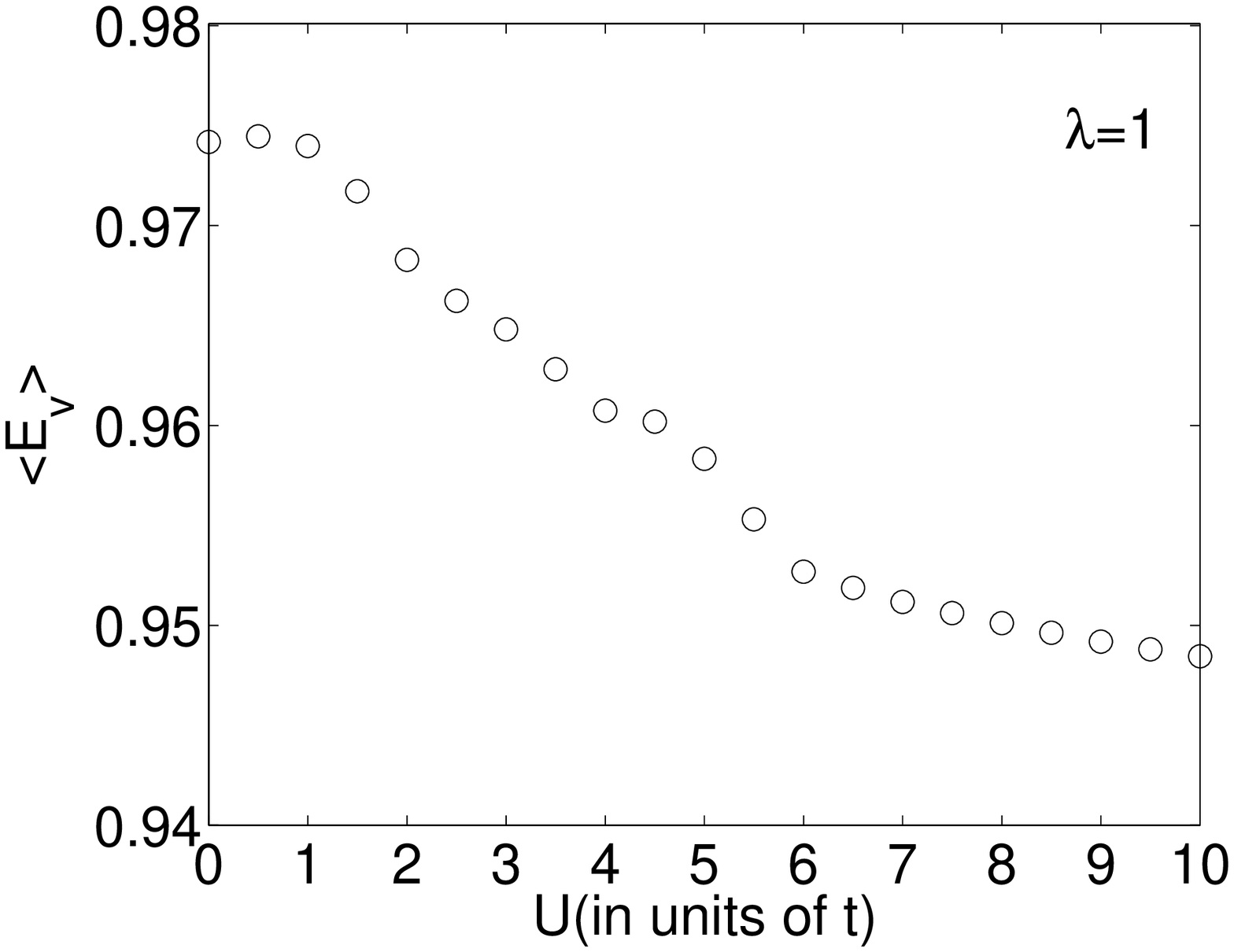}
(b)\includegraphics[width=2.5in]{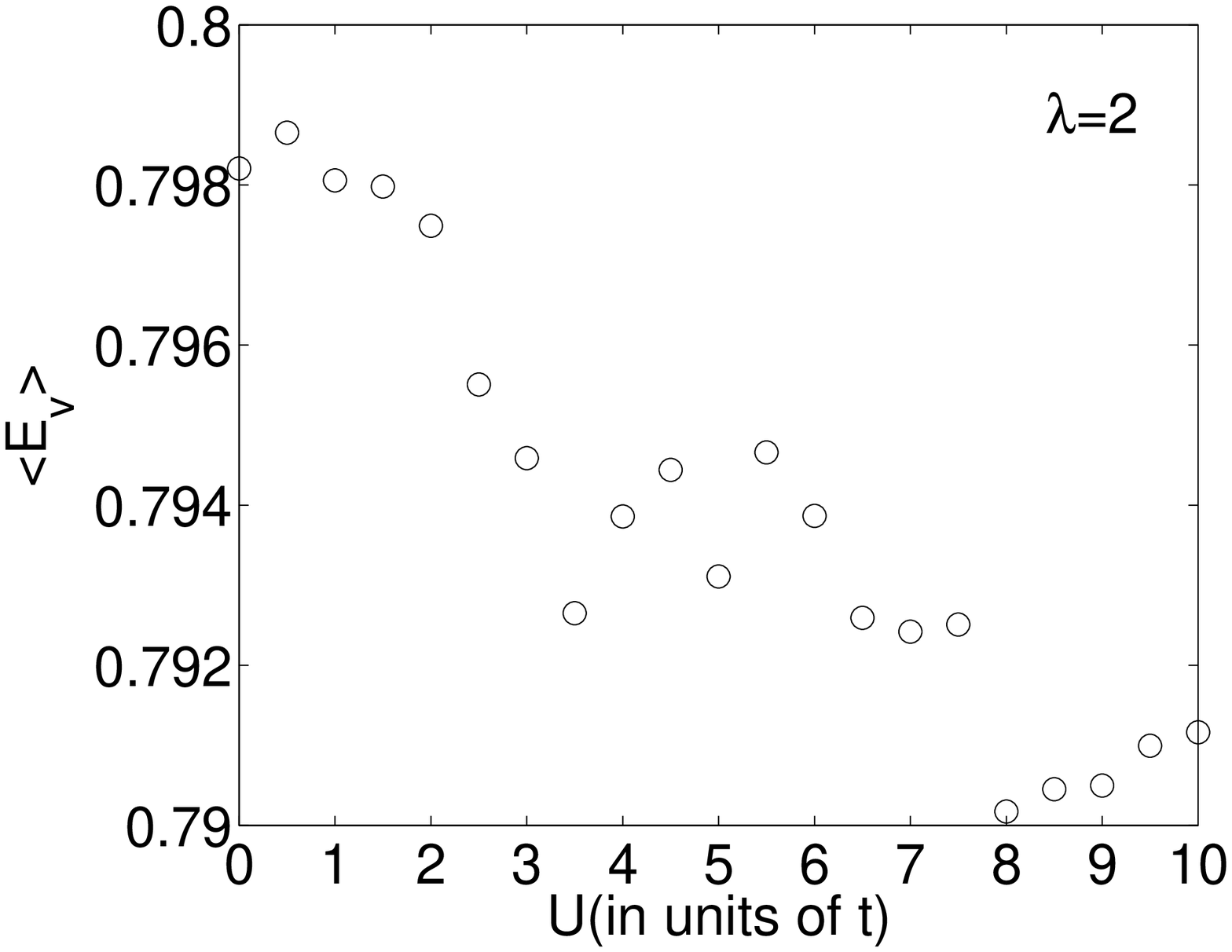}
(c)\includegraphics[width=2.5in]{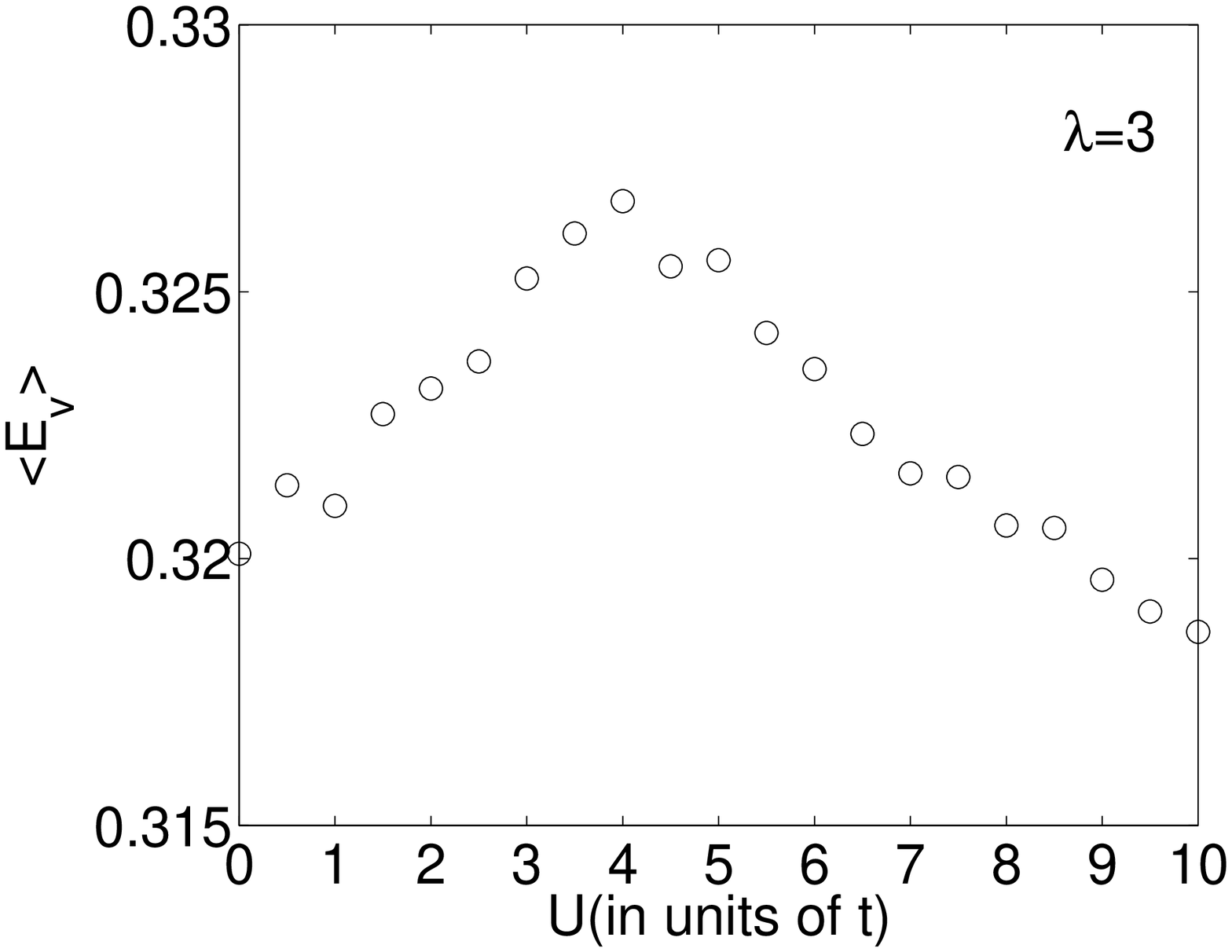} \caption{ The
spectrum-averaged von Neumann entropy $\langle E_v\rangle$ as a
function of interaction strengths $U$ for
(a)$\lambda=1$,(b)$\lambda=2$ and (c)$\lambda=3$, respectively.
Here $N=89$.}\label{Fig6}
\end{figure} %Fig6

For $U>0$, the spectrum-averaged von Neumann entropy $\langle
E_v\rangle$ as a function of $U$ at different $\lambda$ are shown
in Fig.\ref{Fig6}. Here we choose $\lambda=1,2$ and $3$ as
examples for extended, critical and localized regimes,
respectively. At $\lambda=1$ and $\lambda=2$, $\langle E_v\rangle$
monotonically decreases when $U$ increases from zero. At
$\lambda=3$ $\langle E_v\rangle$ increases for small $U$ and
decreases for large $U$, respectively, which is similar to that
shown in Fig.\ref{Fig3} for TIP in a disordered potential chain.
For $\lambda$ in extended and localized regimes,  $\langle
E_v\rangle$ as a function of $U$ are similar to that for
$\lambda=1$ and $\lambda=3$, respectively. We also find that at
the extended and critical regimes, the newly created eigenstates
due to the interaction have small $E^\alpha_v$, which will reduce
the value of $\langle E_v\rangle$. At the localized regime, the
effect of interaction $U$ on $\langle E_v\rangle$ is similar to
that in the disordered potential system, so $\langle E_v\rangle$
increases for small $U$, decreases for large $U$, respectively.

For TIP in the 1D Harper model, Shepelyansky $et$ $al.$
\cite{sh96,ba96,ba97} found that the interaction would induce
localization effect for all $\lambda$ . At the same time,
Evangelou and Katsanos \cite{ev97} found that, in the
extended($\lambda<2$) and the critical($\lambda=2$) regimes, the
velocity and the diffusion coefficient of TIP will decreases due
to the localized pairing states. In the localization
regime($\lambda>2$), they found that propagation enhancement for
small interaction and strong localization for large interaction,
as in disorder systems. Apparently there are discrepancies in
their results at the localization regime. Comparing to our
results, in the extended and the critical regimes, we find
$\langle E_v\rangle$ decreases with $U$, which agrees with both of
their conclusions that the interaction can induce localization
effect.\cite{sh96,ba96,ba97,ev97} In the localized regime, we find
$\langle E_v\rangle$ increases for small $U$ and decreases for
large $U$, respectively, which is consistent with the results of
Evangelou and Katsanos \cite{ev97} that the interaction has
different effects on localization properties at small and large
$U$.

\subsection{TIP in a slowly varying potential chain}
Nest we study TIP moving in a 1D system based on the slowly
varying potential model .\cite{gr88,th88,sa88} From
Eq.(\ref{form1}), the eigenvalue equation  can be described by
\begin{eqnarray}
\lefteqn{[\lambda \cos (\pi \alpha n_1^\upsilon+{\beta}_1
)+\lambda \cos
(\pi \alpha n_2^\upsilon +{\beta}_2) +U\delta_{n_1,n_2}]\psi_{n_1,n_2} }\hspace{1cm}\nonumber\\
& &+t(\psi_{n_1+1, n_2} +\psi_{n_1-1, n_2} +\psi_{n_1, n_2+1}+ \psi_{n_1, n_2-1}) \nonumber\\
& &=E\psi_{n_1, n_2}, \label{form14}
\end{eqnarray} %formu(14)
here $\lambda$, $\alpha$ and $\upsilon $ are positive numbers. For
 $\alpha$ irrational with  $\upsilon\geq2$ or $\upsilon=1$, this is
 equivalent to the models discussed in Sec.~\ref{sec3} (A) and (B),
 respectively. For $0<\upsilon <1$, in the absence
of the interaction U, it is well known \cite{sa88} that there are
two mobility edges at $E_c = \pm (2.0 - \lambda ) $ provided
$\lambda<2.0$. It is found that extended states are in the middle
of the band ( $\left| E \right| < 2.0 - \lambda$ ) and localized
states are at the band edge ( $2.0 - \lambda  < \left| E \right| <
2.0 + \lambda$ ). In other words, for $\lambda<2.0$, the extended
and localized eigenstates coexist in contrast to the models
studied in Sec.~\ref{sec3} (A) and (B). For $\lambda>2.0$, all
states are found to be localized. So in this model there are
always localized eigenstates at $\lambda\neq0$. Obviously the
spectrum properties are different from that for the Anderson model
and the Harper model.
\begin{figure}%[!ht]%Fig7
\includegraphics[width=2.5in]{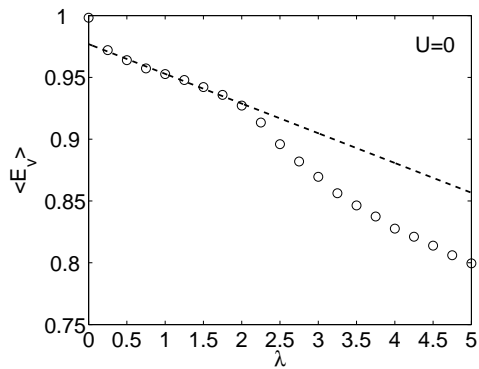}
\caption{ The spectrum-averaged von Neumann entropy $\langle
E_v\rangle$ as a function of potential strength $\lambda$ at
$U=0$. The dashed line is linearly fitted for the corresponding
data at $0<\lambda\leq2$. Here $\pi \alpha=0.2$, $\upsilon=0.7$
and $N=89$.}\label{Fig7}
\end{figure} %Fig7

\begin{figure}%[!ht]%Fig8
\includegraphics[width=2.5in]{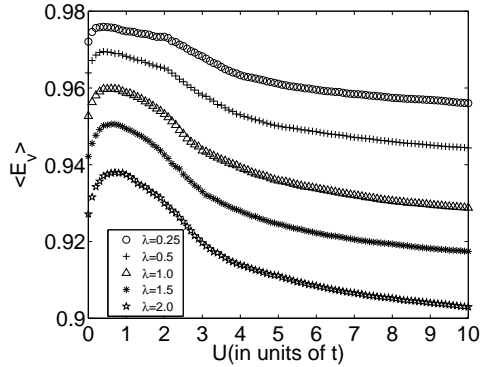}
\caption{ The spectrum-averaged von Neumann entropy $\langle
E_v\rangle$ as functions of interaction strengths $U$ at different
$\lambda$. Here $\pi \alpha=0.2$, $\upsilon=0.7$ and
$N=89$.}\label{Fig8}
\end{figure} %Fig8

Fig.\ref{Fig7} shows the spectrum-averaged von Neumann entropy
$\langle E_v\rangle$ as a function of  $\lambda$ at $U=0$. For
$0<\lambda\leq2$ the data can  be well fitted into a line. For
$\lambda>2$ all data points lie far away from the proposed line,
i.e., there is an abrupt decrease in $\langle E_v\rangle$ at the
critical parameter $\lambda=2$. The critical parameter separates
the localized regime($\lambda>2$) and the regime where the
localized and extended eigenstates coexist($\lambda<2$). This
result is consistent with the spectrum properties for the
model,\cite{sa88} so the spectrum-averaged von Neumann entropy
$\langle E_v\rangle$ can describe the localization properties for
two-particle systems.

Fig.\ref{Fig8} shows the spectrum-averaged von Neumann entropy
$\langle E_v\rangle$ versus the interaction $U$
 at $\lambda\leq2$. The results are similar to
that for $\lambda>2$. We again find that for all $\lambda$, as $U$
increases, $\langle E_v\rangle$ increases for small $U$, decreases
for large $U$, respectively, similar to that shown in
Fig.\ref{Fig3} for TIP in a disordered potential chain.  It
indicates that propagation enhancement for small interaction and
strong localization for large interaction. In other words, as long
as there are localized eigenstate at single-particle models, the
effect of interaction on the localization properties of TIP is
similar to that in disorder systems.
\section{\label{sec4}Conclusions and Discussions}

With the help of von Neumann entropy, we have studied the effect
of the on-site interaction $U$ on the localization properties of
TIP in 1D disordered, quasiperiodic and slowly varying potential
systems, respectively.

%In the case of vanishing interaction($U=0$), there is a strong
%correlation between the spectrum-averaged von Neumann entropy
%$\langle E_v\rangle$ and the single-particle eigenstate
%properties, which means that $\langle E_v\rangle$ is a suitable
%quantity to describe the localization properties for two-particle
%systems.

For TIP in a disordered potential chain and a slowly varying
potential chain , as $U$ increases from zero, we find that at
first $\langle E_v\rangle$ increases for small $U$, then decreases
as $U$ gets large. It means that there are propagation enhancement
for small interaction and stronge localization for large
interaction. For TIP in 1D Harper model with particles in the
extended($\lambda<2$) and critical($\lambda=2$) regimes, $\langle
E_v\rangle$ decreases as $U$ increases, which indicates that the
interaction would induce localization effect, while in the
localized regime($\lambda>2$), the $\langle E_v\rangle$ as a
function of $U$ is similar to that for TIP in the disordered
potential chain. From our studies, we find that $\langle
E_v\rangle$ is a suitable quantity to describe the localization
properties for two-particle systems.

Summarizing all results from the three models, we can conclude
that provided localized eigenstates at single-particle case exist,
the delocaliation (localization) effect happens for small
interactions (large interactions),  while single particle states
are extended or critical, the interaction always induces the
localization effect. According to our results, we propose a
consistent interpretation of the discrepancies between previous
numerical results.

\begin{acknowledgments}
We would like to thank Zhongxia Yang and Shijie Xiong for a
critical reading of the manuscript and useful discussions. This
work is partly supported by National Nature Science Foundation of
China under Grants Nos. 90203009, 10175035, and 10674072, by SRFDP
No.20060319007, by EYTP of MOE, China, by NSF of Jiangsu No.
06KJD140135, and by the Foundation of NJUPT No. NY205050.

%This work is partly supported by the National Nature Science
%Foundation of China under Grants Nos. 90203009, 10175035, and
%10674072, by SRFDP under Grant No.20060319007, by the Excellent
%Young Teacher Program of MOE, China, by the Nature Science
%Foundation of Jiangsu Province Education Department of China under
%Grant No. 06KJD140135, and by the Foundation of Nanjing University
%of posts and telecommunications under Grant No. NY205050, China.
\end{acknowledgments}


\begin{thebibliography}{99}
\bibitem{li68}E. H. Lieb and F. Y. Wu, Phys. Rev. Lett. {\bf 20},
1445 (1968).
\bibitem{an58}P. W. Anderson, Phys. Rev.{\bf 109},
1492 (1958).
\bibitem{kr93}B. Kramer and A. MacKinnon, Rep. Prog. Phys. {\bf 56}, 1469
(1993). %one particle moving in a random potential

\bibitem{le85}P. A. Lee and T. V. Ramakrishnan, Rev. Mod. Phys. {\bf 57}, 287
(1985).
\bibitem{be94}D. Belitz and T. R. Kirkpatrick,
Rev. Mod. Phys. {\bf 66}, 261 (1994). %interaction between the particles \cite{be94}


\bibitem{sh94}D. L. Shepelyansky, Phys. Rev. Lett. {\bf 73}, 2607
(1994). %random

\bibitem{sh96}D. L. Shepelyansky, Phys. Rev. B {\bf 54}, 14896
(1996). %Harper
\bibitem{ev97}S. N. Evangelou, D. E. Katsanos, Phys. Rev. B {\bf 56}, 12797
(1997). %Harper diagonal and dynamical

\bibitem{br06}P. E. de Brito, E. S. Rodrigues,and H. N. Nazareno,
Phys. Rev. B {\bf 73}, 014301 (2006). %aperiodic

\bibitem{ev96}S. N. Evangelou, Shi-Jie Xiong, and E.N. Economou,
Phys. Rev. B {\bf 54}, 8469 (1996). %Diagonal Anderson

\bibitem{op96} F. von Oppen, T. Wettig, and J. M\"{u}ller,
   Phys. Rev. Lett. {\bf 76}, 491
(1996). % Green
\bibitem{so97}P. H. Song and Doochul Kim, Phys. Rev. B {\bf 56}, 12217
(1997). %recursive Green
\bibitem{fr97}K. Frahm, A. M\"{u}ller-Groeling, J.-L. Pichard,
and D. Weinmann, Phys. Rev. Lett. {\bf78}, 4889(1997).

\bibitem{ro97}R. A. R\"{o}mer and M. Schreiber, Phys. Rev. Lett.
    {\bf 78}, 515 (1997). %TMM

\bibitem{ba96}A. Barelli, J. Bellissard, P. Jacquod
 and D. L. Shepelyansky, Phys. Rev. Lett. {\bf 77}, 4752(1996). %butterfly

\bibitem{ba97}A. Barelli, J. Bellissard, P. Jacquod and D. L.
Shepelyansky, Phys. Rev. B {\bf 55}, 9524 (1997). % Hofstadter butterflies
%%%%%%%%%%%%%%%%%%%%%%%%%%%%%%%%%%%%%%%%%%%%%%
\bibitem{bo00}See, for example, The Physics of Quantum Information, edited
by D. Bouwmeester, A. Ekert, and A. Zeilinger (Springer, Berlin,
2000).
%%%%%%5
\bibitem{za02}P. Zanardi, Phys. Rev. A {\bf 65},
  042101 (2002). %Isomorphi & local entangle

\bibitem{yu04}Yu Shi, J. Phys. A {\bf 37}, 6807 (2004). %second quantized

\bibitem{gu04}S.-J. Gu, S.-S. Deng, Y.-Q. Li, and H.-Q. Lin,
Phys. Rev. Lett. {\bf 93}, 086402 (2004).%extended Hubbard model

\bibitem{la06}D. Larsson and H. Johannesson, Phys. Rev. A {\bf73}, 042320
(2006). %Fermions at a Quantum Phase Transition
\bibitem{go06}L. Y. Gong and P. Q. Tong, Phys. Rev. E {\bf74},
056103 (2006).
\bibitem{bu06}F. Buscemi, P. Bordone, and A.
Bertoni, phys. Rev. A {\bf73},
052312 (2006). %Entanglement dynamics of electron-electron scattering

\bibitem{sa88}S. Das Sarma, S. He, and X. C. Xie, Phys. Rev. Lett.
{\bf 61}, 2144 (1988). %slowly varying potential model
%
\bibitem{gr88}M. Griniasty and S. Fishman, Phys. Rev. Lett.
{\bf60}, 1334 (1988).
\bibitem{th88}D. J. Thouless, Phys. Rev. Lett.
{\bf61}, 2141 (1988).%varying potential
\bibitem{go07}For random potentials, we use the following definition
of the spectrum-averaged von Neumann entropy: $\langle E_v \rangle
=\overline{ \frac{1}{M}\sum\limits_{\alpha} {E^\alpha_v
}}=\frac{1}{K}\sum\limits_{k=1}^K(\frac{1}{M}\sum\limits_{\alpha}
{E^\alpha_v })_k$,  here $\overline{X}$ denotes the disorder
average, and $K$ the number samples.

\bibitem{au80}S. Aubry and G. Andr\'{e}, Ann. Isr. Phys.
Soc. {\bf 3}, 133 (1988). %Intensively analytical and numerical studies
\bibitem{so85}J. B. Sokoloff, Phys. Rep. {\bf 126}, 189 (1985).
%Intensively analytical and numerical studies

%\bibitem{ei99}A. Eilmes, U. Grimm, R. A. R\"{o}mer, and M. Schreiber, Eur. Phys. J. B
%{\bf 8},547(1999). %MIT


\end{thebibliography}
\end{document}